\documentclass[a4paper,twocolumn,11pt,unpublished,accepted=2020-02-25]{quantumarticle}
\pdfoutput=1

\usepackage[utf8]{inputenc}
\usepackage[english]{babel}
\usepackage{amsmath}
\usepackage{hyperref}
\usepackage{graphicx}
\usepackage{sistyle}
\usepackage{todonotes}
\usepackage{gensymb}
\usepackage{bbold}
\usepackage{bm}
\usepackage{amsfonts}
\usepackage{dsfont}
\usepackage{amssymb}
\usepackage{latexsym}
\usepackage{fancyhdr}
\usepackage[bbgreekl]{mathbbol}
\usepackage{epsfig}
\usepackage{color}
\usepackage{textcomp}
\usepackage{mathrsfs}
\usepackage{ulem}

\usepackage{tcolorbox}
\tcbuselibrary{theorems}

\newcommand{\ket}[1]{| #1 \rangle}

\newcommand{\prjct}[1]{|#1 \rangle\!\langle #1 |}

\newcommand{\cI}{{\mathcal{I}}}

\newcommand{\tr}{\mathrm{Tr}}

\newcommand{\id}{\mathbb{1}}
\newcommand{\mapid}{\text{id}}
\newcommand{\tot}{\!\otimes\!}
\newcommand{\cB}{\mathcal{B}}

\newcommand{\cF}{\mathcal{F}}
\newcommand{\cH}{\mathcal{H}}
\newcommand{\cM}{\mathcal{M}}
\newcommand{\sA}{\mathscr{A}}
\newcommand{\sB}{\mathscr{B}}
\newcommand{\dC}{\mathds{C}}

\newcommand{\be}{\begin{equation}} 
\newcommand{\ee}{\end{equation}}

\begin{document}

\title{Device-independent characterization of quantum instruments}

\author{Sebastian Wagner}
\affiliation{Department of Physics, University of Basel, Klingelbergstrasse 82, CH-4056 Basel, Switzerland}
\author{Jean-Daniel Bancal}
\affiliation{Department of Physics, University of Basel, Klingelbergstrasse 82, CH-4056 Basel, Switzerland}
\author{Nicolas Sangouard}
\affiliation{Department of Physics, University of Basel, Klingelbergstrasse 82, CH-4056 Basel, Switzerland}
\author{Pavel Sekatski}
\affiliation{Department of Physics, University of Basel, Klingelbergstrasse 82, CH-4056 Basel, Switzerland}
%\date{\today}

\begin{abstract}
Among certification techniques, those based on the violation of Bell inequalities are appealing because they do not require assumptions on the underlying Hilbert space dimension and on the accuracy of calibration methods. Such device-independent techniques have been proposed to certify the quality of entangled states, unitary operations, projective measurements following von Neumann's model and rank-one positive-operator-valued measures (POVM). Here, we show that they can be extended to the characterization of quantum instruments with post-measurement states that are not fully determined by the Kraus operators but also depend on input states. We provide concrete certification recipes that are robust to noise. \\
\end{abstract}
%\pacs{03.65.Ud, 43.40.Dx}
\maketitle

\section*{Introduction}
Experiments using either NV centers~\cite{Hensen15}, photon pair sources~\cite{Shalm15, Giustina15} or neutral atoms~\cite{Rosenfeld17} have recently been used to test Bell inequalities~\cite{Bell64} in a very convincing way. The observed Bell inequality violations have brought new and fascinating insights about nature by showing that some correlations cannot be explained by locally causal models. These experiments also revolutionize branches of applied physics like randomness generation~\cite{Colbeck09, Pironio10, Christensen13, Yang18, Yang18bis, Bierhorst18, Shen18}
by making it device-independent, i.e. the randomness guarantees hold without assumptions on the underlying Hilbert space dimension and on the accuracy of calibration methods.\\

The possibility of randomness generation from Bell inequalities is clear when one realizes  that the only situation allowing for a maximal quantum violation of the simplest Bell inequality \cite{CHSH69}, within the quantum formalism, consists in using  complementary Pauli measurements on a maximally-entangled two-qubit state~\cite{Popescu92,McKague12}. This means that the violation of a Bell inequality can certify quantum states and von Neumann measurements directly, without resorting to tomography. Mayers and Yao were among the very first ones to highlight the usefulness of Bell tests as characterization methods, a technique that they called self-testing~\cite{Mayers04}. Self-testing has been applied to many entangled states~\cite{Mayers04, McKague13, Coladangelo16, McKague12, Wu14}, projective measurements~\cite{Mayers04, Bancal15, Chen16, Cavalcanti16, Kaniewski17, Bowles18, Renou18, Bancal18}, and unitary operations~\cite{Sekatski18}. \\

Efforts are being devoted to characterise measurements not captured by the usual von Neumann model. Refs.~\cite{Gomez16,Smania18} for example, showed how to characterise rank-one POVMs that are not composed of orthogonal projection. Less is known for measurements whose post-measurement state is not fully determined by the Kraus operator associated with the measurement result, but also depends on the input state. In this case, the measurement statistics and the post-measurement states have to be considered together in order to verify that a measurement achieves the ideal trade-off between disturbance and information gain. Such quantum instruments~\cite{davies70}, sometimes called weak measurements~\cite{Gross18}, can be more efficient in practice than projective or rank-one measurements e.g. for generating randomness. Whereas randomness generation based on projective measurements requires at least as many maximally-entangled states as the number of certified random bits, an arbitrary number of random bits can in principle be extracted from a single maximally-entangled state by applying successive quantum instruments which do not break entanglement~\cite{Silva15, Curchod17, Curchod18, Coyle18}. The certification of such measurements is thus not only of fundamental interest but could be used in practice to characterise the potential of an actual quantum instrument for producing large amounts of device-independent randomness with a single entangled state. \\

In this manuscript, we provide a recipe to certify quantum instruments that are neither projective measurements nor rank-one POVMs, i.e. we certify the states conditioned on each outcome as well as the probability of each outcome.
We also derive a new class of Bell inequalities suitable for the robust self-testing of partially-entangled two-qubit states. Our final recipe is realistically robust to experimental noise.\\

\section*{Device-independent certification of quantum instruments}
We consider an unknown instrument, that we denote $\mathcal{M},$ and our goal is to show that it behaves as an ideal instrument $\mathcal{\overline M}$ in a device-independent way. We start by clarifying the formulation, then describe the proposed recipe before discussing the results.

\subsection*{Formulation}
Consider  an ideal noise-free quantum instrument $\mathcal{\overline M}$ with $k$ outcomes operating on a single system of dimension $d$. It is represented by a collection of $k$ Kraus operators $\{\overline K_\ell\}_{\ell=0}^{k-1}$ satisfying the completeness relation $\sum_\ell \overline K_\ell^\dag \overline K_\ell = \id.$ Each Kraus operator defines a completely positive map $\rho \mapsto \overline K_\ell \rho \overline K_\ell^\dag$.
Given a state $\rho_\sB \in L(\mathds{C}^d),$ the probability to observe the outcome $\ell$ is given by the Born rule $\overline p_\ell=\tr{\left(\overline K_\ell \rho_\sB \overline K_\ell^\dag\right)}$ and the normalized post-measurement state for this outcome is $\varrho_\ell=\frac{1}{\overline p_\ell}\overline K_\ell \rho_\sB \overline K_\ell^\dag$. Note that the dimension of $\varrho_\ell$ is not necessarily the same than the one of $\rho_\sB.$ To illustrate our method, we consider in this work measurements operating on qubits and returning qubits, cf. Eq. \eqref{def_ref_measurement}. \\

Von Neumann measurements and rank-one POVMs correspond to the specific case where $\{\overline K_\ell\}$ are proportional to projectors onto pure states $|\psi_\ell\rangle$, that is $\overline K_\ell=\eta_\ell \, \prjct{\psi_\ell}$.
In this case, the post-measurement state corresponding to the outcome $\ell$ is $|\psi_\ell\rangle$ independently of the pre-measured state of the system. 
Performing such a measurement on a physical system can extract full information about its state but also disturbs it maximally, e.g. it breaks all entanglement the system might have with the rest of the world. \\

Quantum instruments that are neither Von Neumann measurements nor rank-one POVMs introduce less disturbance in the system at the price of extracting less information~\cite{Banaszek01}. This has the benefit of preserving interesting and useful features such as entanglement while nevertheless revealing information about the system. As an example, for $\theta\in[0,\pi/4]$ the Kraus operators
\begin{align}
\label{Kraus0}
&\overline K_0(\theta) = \cos (\theta)  \prjct{0} + \sin (\theta)  \prjct{1},\\
\label{Kraus1}
&\overline K_1(\theta) = \sin(\theta) \prjct{0} + \cos (\theta) \prjct{1}
\end{align}
are associated to quantum instruments which tend to be projective in the limit $\theta \rightarrow 0,$ and the identity in the limit when $\theta \rightarrow \pi/4$. Such an instrument is sufficient to implement the scheme proposed in~\cite{Curchod17, Curchod18, Coyle18} to produce more randomness than possible with von Neumann measurements.  \\

Whether the considered measurement is a von Neumann measurement or not, it can be fully characterised by the map
\begin{align}
\label{def_ref_measurement}
\overline{\mathcal{M}} :  \,  L(\mathds{C}^2) &\rightarrow L(\mathds{C}^2 \otimes \cH_R) \ ,\\
\nonumber
\sigma &\mapsto \sum_\ell  \left( \overline K_\ell\, \sigma\, \overline K_\ell^\dag\right) \otimes \prjct{\ell}_R \ ,
\end{align}
where we have introduced a register $R$ indicating the outcome. In comparison to the map generated by a single Kraus operator $\overline K_\ell$, the map $\overline \cM$ is trace-preserving by construction, hence defines a quantum channel. We note that every set of Kraus operators uniquely defines a map through Eq.~\eqref{def_ref_measurement}, and any map of this form uniquely defines the post-measurement states, and hence corresponds to a quantum instrument. We illustrate our method by showing how to certify a quantum instrument of the form given in Eq. \eqref{def_ref_measurement} with the Kraus operators written in Eqs. \eqref{Kraus0} and \eqref{Kraus1}.\\

\begin{figure}
\centering\includegraphics[width=0.7\columnwidth]{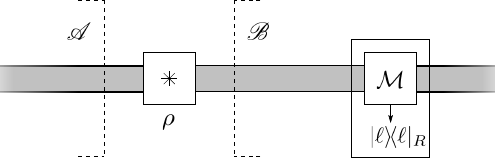}
\caption{Scheme of the actual experiment that is used to characterise the measurement box $\mathcal{M}$ in a device-independent way. The source which is represented by a box with a star, produces an unknown bipartite state $\rho$ shared between $\sA$ and $\sB.$ Party $\sB$ performs the measurement. The pre- and post-measurement states can be measured with additional measurements named $A_{0/1}$ for party $\sA$ and $B_{0/1/2/3}$ for party $\sB$ that are also unknown (not represented).}
\label{Fig1}
\end{figure}

\begin{figure}
\centering\includegraphics[width=0.8\columnwidth]{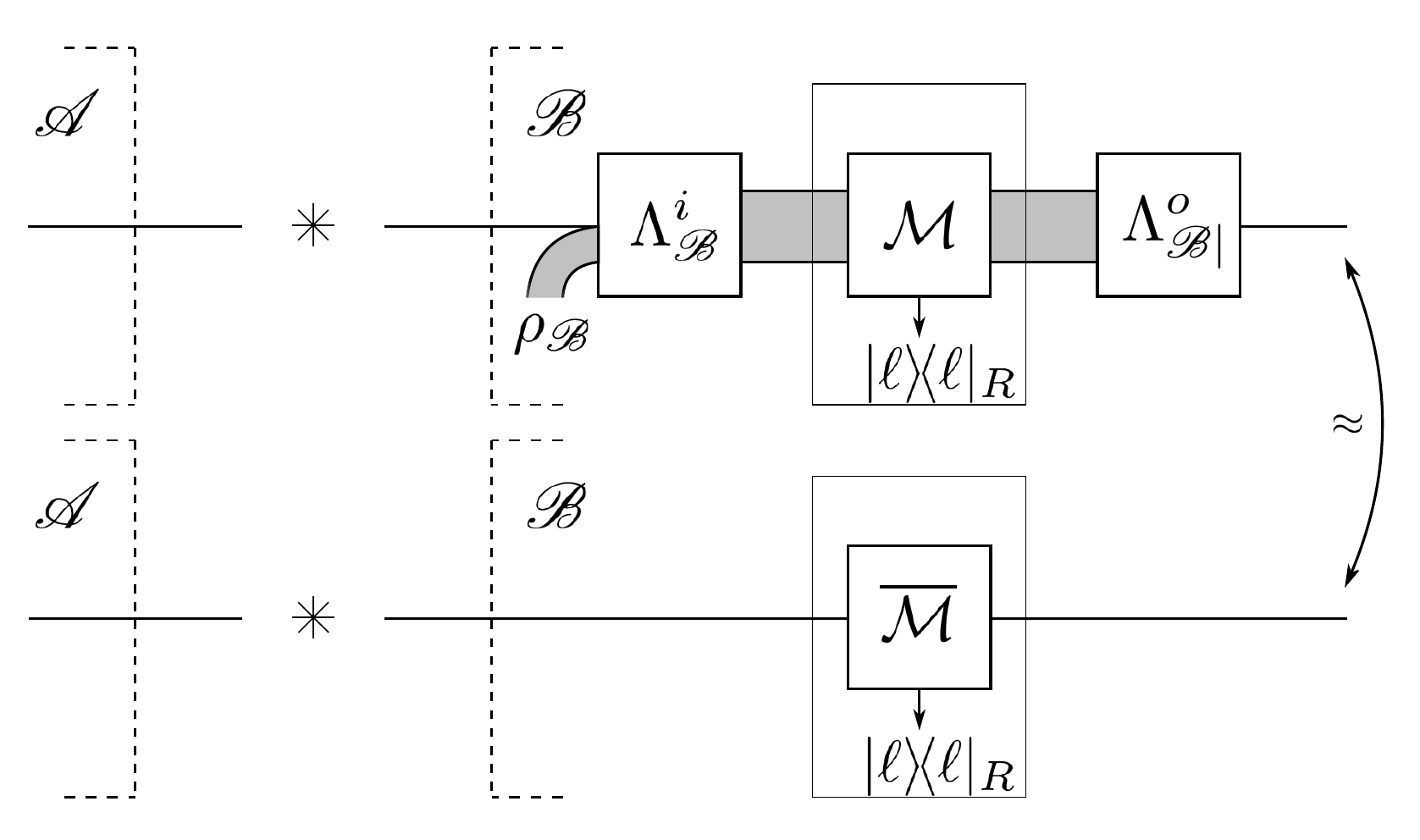}
\caption{To characterise an unknown measurement $\mathcal{M},$ we compare the action of this black box supplemented with injection
maps $\Lambda_\sB^i$ and $\Lambda_{\sB|}^o$ with the action of a reference measurement $\overline{\mathcal{M}}$ on one half of a maximally
entangled two-qubit state $\ket{\phi^+}.$}
\label{Fig2}
\end{figure}

An actual - possibly noisy - measurement (see Fig. \ref{Fig1}) acting on a Hilbert space $\cH_\sB$ can be described by the following map
\begin{align}
\label{def_unknown_measurement}
\mathcal{M} :  \,  L(\cH_\sB) &\rightarrow L(\cH_\sB \otimes \cH_R) \ , \\
\nonumber 
\sigma &\mapsto \sum_\ell M_\ell[\sigma] \otimes \prjct{\ell}_R \ ,
\end{align}
where $M_\ell$ are the completely positive maps associated to the outcomes $\ell$. In general, these maps may not be expressed in terms of a single Kraus operator, but as a combination of several ones, i.e. $M_\ell[\sigma] = \sum_m K_{\ell,m} \sigma K_{\ell,m}^\dag$, with $\sum_{\ell,m} K_{\ell,m}^\dag K_{\ell,m}=\id$. In order to show that a measurement described by $\cM$ acts like a target measurement $\overline\cM$, it is sufficient to identify a subspace of $\cH_\sB$ on which the action of $\cM$ is similar to the one of $\overline\cM$. Moreover, a map is fully described by its action on one half of a maximally-entangled state, a result known as the Choi-Jamiolkowski isomorphism~\cite{Choi75,Jamiolkowski72}. Therefore, the equivalence between the considered measurement and the target one is obtained by showing that there exist completely positive trace-preserving maps
\begin{align}
\Lambda_\sB^i : L(\mathds{C}^2) \to L(\cH_\sB)
\end{align}
and
\begin{align}
\label{eq: Lout}
\Lambda_\sB^o : L(\cH_\sB \otimes \cH_R) &\to L(\mathds{C}^2\otimes \cH_R) \ , \\
\nonumber \sigma \otimes \prjct{\ell}_R &\mapsto \Lambda^{o}_{\sB|}[\sigma] \otimes \prjct{\ell}_R \ ,
\end{align}
such that the composition of these maps with the actual measurement is identical to the reference measurement, that is
\begin{equation}\label{eq:KKbar}
(\mapid \otimes \Lambda_\sB^o\!\circ\! \cM \!\circ\!\Lambda_\sB^i)[\prjct{\phi^+}] = (\mapid\otimes\overline\cM)[\prjct{\phi^+}].
\end{equation}
The injection map $\Lambda^i_\sB$ and the output map $\Lambda^o_\sB$ identify subspaces and subsystems in which the measurement $\mathcal{M}$ acts as the reference measurement $\overline{\mathcal{M}},$ see Fig. \ref{Fig2}. Note that
the output map does not depend on the measurement output, cf discussion below.
Also note that since all possible outcomes appear in the definition of the maps $\overline\cM$ and $\cM$ -- the corresponding Choi states include a description of the labels -- equality~\eqref{eq:KKbar} guarantees at the same time that the outcome states are as expected \textit{and} that each outcome appears with the desired probability.\\

The previous equality cannot be satisfied in an actual experiment due to unavoidable imperfections. Following~\cite{Raginsky01,Belavkin05}, we thus propose an extension for quantifying the distance $\cF(\mathcal{M},\overline{\mathcal{M}}) $ between $\mathcal{M}$ and $\overline{\mathcal{M}}$ using
\begin{align}
\label{Definition_certificationGM}
\cF(\mathcal{M},\overline{\mathcal{M}}) = &\\
\nonumber
\underset{\Lambda_\sB^i, \Lambda_\sB^o}{\max} F\Big(&\!\left(\mapid\otimes\Lambda_\sB^o \!\circ\! \mathcal{M} \!\circ\! \Lambda_\sB^i\right) [\prjct{\phi^+}] ,\\
& (\mapid\otimes\overline{\mathcal{M}})[\prjct{\phi^+}]\Big) ,\nonumber
\end{align}
where $F(\rho,\sigma)=\tr{\!\sqrt{\!\!\sqrt{\rho}\,\sigma\!\sqrt{\rho}}}$ is the Uhlmann fidelity between two states $\rho$ and $\sigma$.\\

\subsection*{Recipe}
\begin{figure}
\centering\includegraphics[width=0.7\columnwidth]{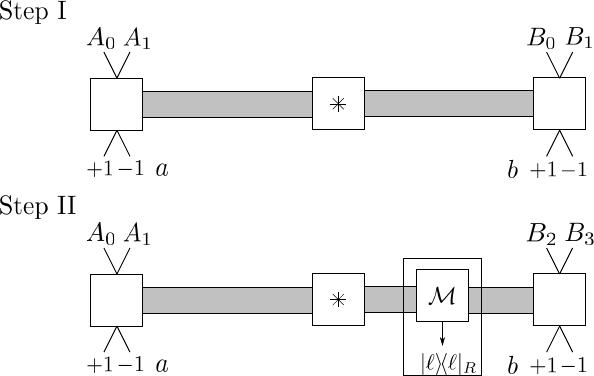}
\caption{Recipe for bounding the quality of the measurement box $\mathcal{M}$ in 2 steps. Step I is used to characterize the state of the source while Step II gives a certificate of the post-measurement states. The statistics recorded in each step is then used to certify the quality of measurement $\mathcal M$ device-independently.}
\label{Fig3}
\end{figure}

The aim of this section is to show how the quantity  \eqref{Definition_certificationGM} can be lower bounded in the setup presented in Fig. \ref{Fig3}. In addition to the source producing the bipartite state $\rho$ and the measurement $\mathcal{M}$ to be characterised, each party has a measurement box. The box of party $\sA$ has two inputs $A_0$ and $A_1$ while the one of party $\sB$ has four inputs $B_0 ,B_1, B_2, B_3.$ For each measurement input, a binary outcome is obtained called $a$ for $\sA$ and $b$ for $\sB,$ with $a,b=\pm1.$ The measurement $\mathcal{M}$ can be applied by party $\sB$ before the measurement input is chosen. Although there is no assumption about the Hilbert space dimension and on the proper calibration of the measurement devices, $\mathcal{M}$ can be characterised in two steps: Step I identifies the quality of the state produced by the source while step II is used to characterise the states after each outcome of the measurement to be certified. The certifications associated to steps I and II are then combined to bound $\cF(\mathcal{M},\overline{\mathcal{M}})$ as defined in Eq. \eqref{Definition_certificationGM}  \\

Let us first focus on step I. In this step, the measurement settings $A_{0/1}$ and $B_{0/1}$ are chosen freely and applied directly on the state produced by the source $\rho$ so that party $\sA$ and $\sB$ can estimate the Clauser, Horne, Shimony, and Holt (CHSH) value~\cite{CHSH69} 
\begin{align}
\beta = \sum_{k,j=0}^1 (-1)^{k\cdot j} \langle A_k B_j \rangle .
\end{align}
Here, $\langle A_k B_j \rangle=\sum_{a,b}\,a\,b\, P(a,b|A_k,B_j)$ is the expectation value of measurements $A_k$ and $B_j$. The CHSH value allows one to bound the fidelity of $\rho$ with a maximally-entangled two-qubit state. In particular, the results of Ref.~\cite{Kaniewski16} show that there exist local extraction maps $\Lambda_{\sA}: L(\cH_{\sA}) \to L(\dC^2)$ and $\tilde{\Lambda}_{\sB|}^i: L(\cH_{\sB}) \to L(\dC^2)$  such that 
\begin{align}
\label{CHSH_input}
&F\big((\Lambda_{\sA} \otimes \tilde{\Lambda}_{\sB|}^i)[\rho], \ket{\phi^+}\Big) \\
&\geq F^i = \sqrt{\frac12+\frac12\cdot\frac{\beta-\beta^*}{2\sqrt 2 -\beta^*}}, \nonumber
\end{align}
where $\beta^*=\frac{2(8 + 7\sqrt{2})}{17} \approx 2.11$. Whenever $\beta=2\sqrt{2},$ the formula \eqref{CHSH_input} certifies that the source produces $|\phi^+\rangle$  up to local maps, these maps being explicitly defined from the quantum description of the measurement inputs $A_{0/1}$ and $B_{0/1}.$ In this case, the same maps can be used to show that $\sA$'s settings correspond to the Pauli measurements $\overline A_0=\sigma_z$ and $\overline A_1 = \sigma_x$ while $\sB$'s settings correspond to $\overline B_{0/1} = \frac1{\sqrt{2}}(\sigma_z \pm\sigma_x)$ \footnote{Mathematically this means, for example for Bob's first measurement $B_0$, that there exists maps $\Lambda_\sB^o$ and $\Lambda_\sB^i$ such that Eq.~\eqref{eq:KKbar} holds with $\cM[\sigma]=\sum_{\ell=0}^1 \frac{\id + (-1)^\ell B_0}{2}\sigma\frac{\id + (-1)^\ell B_0}{2}\otimes\prjct{\ell}$ and $\overline\cM[\sigma]=\sum_{\ell=0}^1 \frac{\id + (-1)^\ell \overline B_0}{2}\sigma\frac{\id + (-1)^\ell \overline B_0}{2}\otimes\prjct{\ell}$. Here,$\Lambda_\sB^i$ can be obtained from $\tilde\Lambda_{\sB|}^{i}$ through Proposition 4 of~\cite{Sekatski18} (in that work the roles of Alice and Bob are exchanged).}.\\

\begin{figure}
\centering
\includegraphics[width=0.92\linewidth]{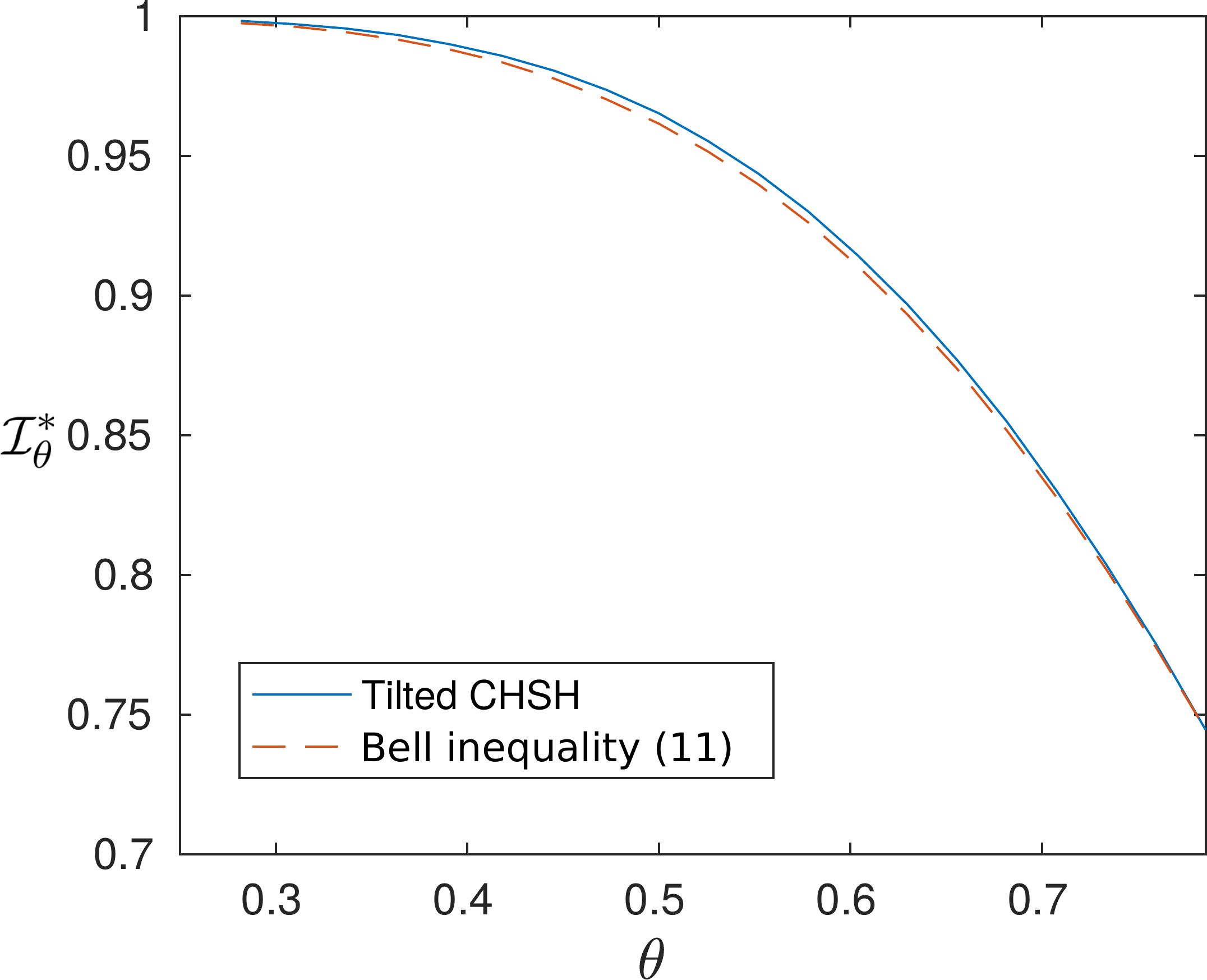}
\caption{Plot of the cutoff parameter $\cI_\theta^*$ as a function of $\theta,$ see Eq. \eqref{output}. The red dashed line is below the blue solid one, indicating that our new inequality achieves non-trivial fidelities for smaller violations compared to the tilted-CHSH inequality. Thus it provides tighter self-testing bounds for partially-entangled states.}
\label{fig:BetaStar}
\end{figure}

In step II, party $\sB$ applies $\mathcal{M}.$ Let us first consider the state conditioned on the outcome $0$, 
\begin{equation}
\nonumber
\varrho_0= \frac{M_0[\rho]}{\tr{\left(M_0[\rho]\right)}} , 
\end{equation}
which is characterized using $A_{0/1}$ and $B_{2/3}.$ In particular, parties $\sA$ and $\sB$ are interested in the Bell inequality 
\begin{align}
\cI_\theta&= \frac14\left[\frac{\langle A_0(B_2-B_3)\rangle}{\sin(b_\theta)}+\frac{\sin(2\theta)}{\cos(b_\theta)}\langle A_1(B_2+B_3)\rangle\right.\nonumber\\
&  \ \ \ \ \ \ \ \ \ \ \ +\left.\cos(2\theta)\left(\langle A_0 \rangle + \frac{\langle B_2-B_3 \rangle}{2\sin(b_\theta)}\right)\right] \nonumber \\
& \leq \frac14\left[\cos(2\theta)+(2+\cos(2\theta))\sqrt{\frac{7-\cos(4\theta)}{5+\cos(4\theta)}}\right]
\label{Bell_Ineq}
\end{align}
with $b_\theta=\arctan\sqrt{(1+\frac{1}{2} \cos^2(2\theta))/\sin^2(2\theta)}$, 
whose maximal quantum value is one by construction. We derived this Bell inequality using the variational method presented in~\cite{Sekatski18,inprep} to self-test partially-entangled two-qubit pure states in a particularly robust manner. Note that for $\theta=\frac\pi4$, Ineq.~\eqref{Bell_Ineq} is the re-normalized CHSH inequality. However, we emphasize that Ineq.~\eqref{Bell_Ineq} is not equivalent to the tilted-CHSH inequality of Refs.~\cite{Acin12, Bamps15} but was carefully constructed for the demands of self-testing presented here. The knowledge of $\cI_\theta$ allows one to bound the fidelity of the conditional state $\varrho_0$, that is, to guarantee the existence of local maps $\Lambda_{\sA}: L(\cH_{\sA}) \to L(\dC^2)$ and  $\Lambda_{\sB|}^{o,\theta}: L(\cH_{\sB}) \to L(\dC^2)$ such that
\begin{align}
&F((\Lambda_\sA \!\otimes\! \Lambda_{\sB|}^{o,\theta})[\varrho_0],\ket{\phi_\theta^0}) \label{output} \\
&\geq F^0_\theta=\sqrt{\cos^2(\theta) + (1-\cos^2(\theta))\frac{\cI_\theta-\cI_\theta^*}{1-\cI_\theta^*}} . \nonumber
\end{align}
Here, $\cI_\theta^*$ is a cutoff parameter corresponding to the violation for which the fidelity matches the square of the largest Schmidt coefficient of $\ket{\phi_\theta^0}=\cos(\theta)|00\rangle+\sin(\theta)|11\rangle$. Fig.~\ref{fig:BetaStar} shows the critical value $\cI_\theta^*$ for both this inequality and the tilted CHSH inequality, as calculated in the Appendix. The previous bound shows that whenever $\cI_\theta$ reaches its maximal quantum value $\cI_\theta=1,$ the state conditioned on the outcome $0$ corresponds to the state $\ket{\phi_\theta^0}$ up to local maps, these maps being explicitly defined from the quantum description of the measurements performed by $A_{0/1}$ and $B_{2/3}$ respectively. The same maps can also be used to show that when $\cI_\theta=1,$ $\sA$'s settings correspond to the Pauli measurements $\overline A_0=\sigma_z$ and $\overline A_1 = \sigma_x$ while $\sB$'s inputs correspond to $\overline B_{2/3} = \cos(b_\theta)\sigma_z \pm \sin(b_\theta) \sigma_x$.\\

The post-measurement state corresponding to the outcome $1$ can be characterized with the same measurement boxes $A_{0/1}$ and $B_{2/3}$, as well as the same local extraction maps $\Lambda_\sA$ and $\Lambda_{\sB|}^{o,\theta}$, see Appendix. Moreover, by including the classical output of the measurement in a global state including a register, as mentioned before, the overall post-measurement state can be written in a compact form
\begin{equation}
\varrho= \sum_{\ell=0}^1 p_\ell \varrho_\ell \otimes \prjct\ell_R \ ,
\end{equation}
with $\varrho_\ell = \frac{M_\ell[\rho]}{\tr{\left(M_\ell[\rho]\right)}}$ the post-measurement state associated to outcome $\ell$, and $p_\ell$ the probability of this outcome. As the certificates for the two branches $\varrho_0$ and $\varrho_1$ are obtained with the same isometries, they can be combined into a single certificate for $\varrho$. In particular, using the orthogonality of the register states, we have
\begin{align}
\nonumber
&F((\Lambda_\sA\otimes\Lambda_\sB^{o,\theta})[\varrho],\sum_{\ell=0}^1 \frac12 \prjct{\phi_\theta^\ell} \otimes \prjct\ell_R) \\
&\geq \sum_\ell \sqrt{\frac{p_\ell}{2}} F^\ell_\theta =: F^o_\theta.
\label{F global}
\end{align}

\begin{figure}
\centering
\includegraphics[width=0.7\linewidth]{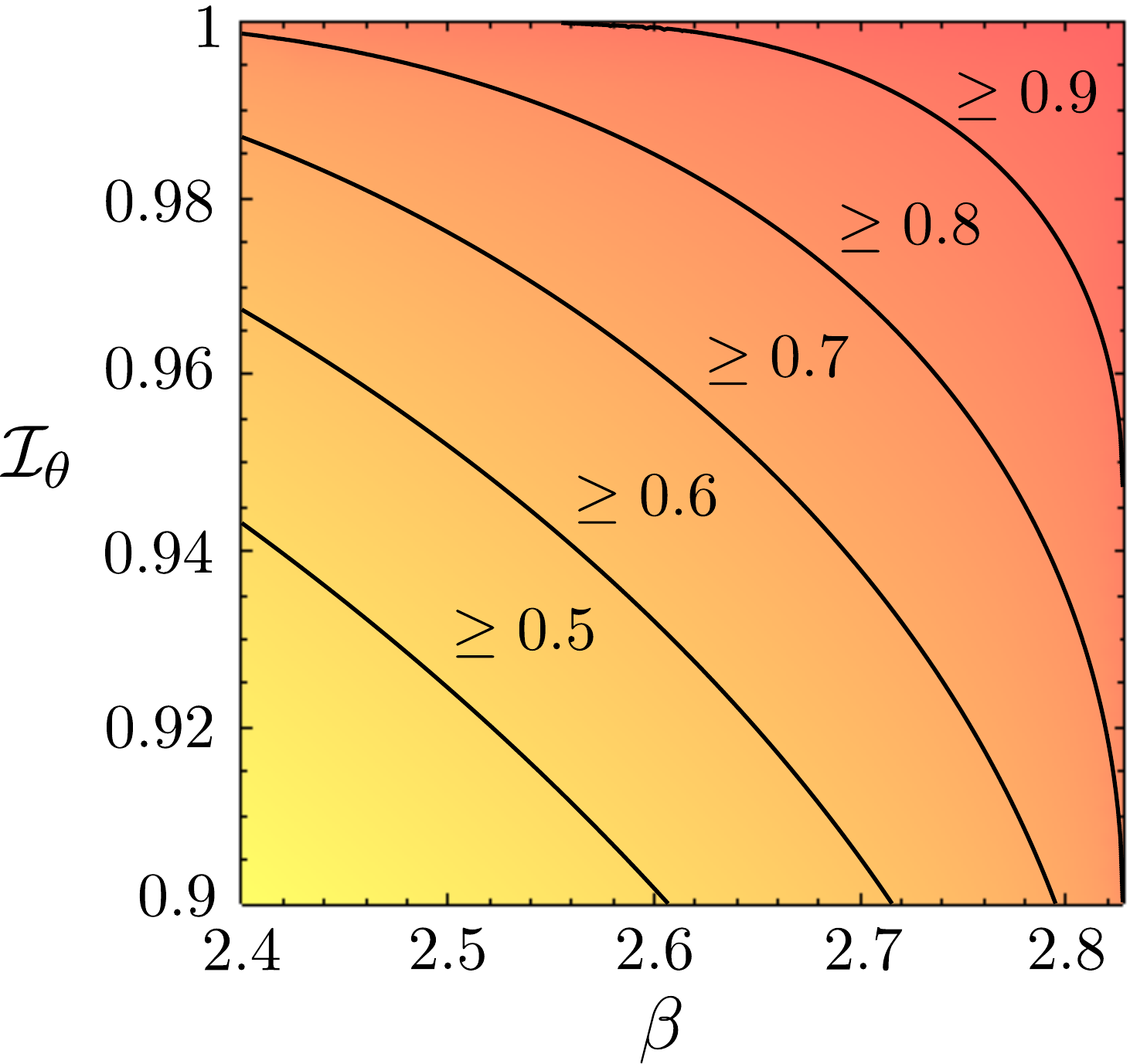}
\caption{Lower bounds on the fidelity $\cF(\mathcal{M},\overline{\mathcal{M}})$ of an unknown measurement $\mathcal{M}$ defined in Eq.~\eqref{def_unknown_measurement} with respect to a reference measurement $\overline{\mathcal{M}}$ defined in Eq.~\eqref{def_ref_measurement} with the Kraus operators given in Eqs.~\eqref{Kraus0}-\eqref{Kraus1} for given CHSH violations $\beta$ and violations $\cI_\theta$ of the Bell inequality~\eqref{Bell_Ineq}; we assume that the second output state appears with probability $p_0=p_1=\frac12$ achieving the violation $\cI_\theta^0,\cI_\theta^1\geq I_\theta$. The plot is for $\theta=(2\pi+7)/22\approx0.6$.}
\label{fig:Results}
\end{figure}

Taking into account the fact that the fidelity cannot decrease under completely-positive trace-preserving maps and using the triangle inequality
$\arccos(F(\rho_1,\rho_3)) \leq \arccos(F(\rho_1,\rho_2)) + \arccos(F(\rho_2,\rho_3))),$
we can prove that the fidelity of the state before and after the measurement can be combined to bound the fidelity of the measurement itself, that is,
\begin{align}
\cF(\mathcal{M},\overline{\mathcal{M}}) \geq \cos(\arccos(F^i)+\arccos(F_\theta^o)) \ ,
\label{eq:FidLowBound}
\end{align}
(see Proposition~5 of \cite{Sekatski18}). Whenever $\beta=2\sqrt{2}$ and $\cI_\theta=1$ for both output states, this bound guarantees that $\cF(\mathcal{M},\overline{\mathcal{M}})=1.$ In noisy scenarios, the fidelity that can be certified is shown in Fig. \ref{fig:Results}. \\

Note that in case where the fidelity of the state before the measurement cannot be assessed, the quality of the measurement cannot be certified. Indeed, if the source produces the state $\ket{\phi_\theta^0}\ket{0}_{B'}+\ket{\phi_\theta^1}\ket{1}_{B'}$, and the measurement simply reads out the auxiliary $B'$ system, then all post-measurement statistics are reproduced. Hence, it is necessary to be able to estimate the quality of the pre-measurement state to give a certificate for the proper functioning of the measurement itself.\\

\subsection*{Discussion}
So far our exposition was focused on the particular example of Kraus operators given by Eq.~\eqref{Kraus0} and \eqref{Kraus1}. Nevertheless, the reader might have noticed that only the tests used to bound the fidelity of the initial state $F^i$ and of the post-measurement states $F^{o}_\ell$ are specific to this example. Consider now a general quantum instrument and assume that one can obtain certificates for the state produced by the source as in Eq.~\eqref{CHSH_input} and for all the post-measurement states as in Eq.~\eqref{output}. If the extraction maps $\Lambda_\sA$ on Alice's side are the same for all certificates, straightforward application of Eqs.~\eqref{F global} and \eqref{eq:FidLowBound} provides a self-testing of the whole instrument. In summary, our approach applies straightforwardly to quantum instruments with an arbitrary number of classical outcomes, including qubit measurements with more than two Kraus operators and measurements on larger-dimensional systems, as well as multipartite instruments. \\

Note also that we focused on a formulation where the extraction maps $\Lambda_{\sB|}^{o}$ in Eq.~\eqref{eq: Lout} act trivially on the classical measurement output $R$, that is the extraction maps applied on the quantum output is independent of the measurement result $\ell$. We believe that this corresponds to the spirit of self-testing in which states and channels are certified up to passive transformations. Nevertheless, a formulation where the extraction maps $\Lambda_{\sB|}^{o, \ell}$ acting on the post-measurement quantum state depend on the measurement outputs $\ell$ is also possible.  In this case it is sufficient to certify the initial and all post-measurement states independently in order to obtain a certificate for the whole instrument. Our derivation then holds as long as Alice's maps are the same.\\

\section*{Conclusion}
We provided a family of Bell inequalities that can be used to self-test non-maximally-entangled two-qubit states with a greater resistance to noise than known Bell inequalities. These results allowed us to derive robust bounds that can be used in practice to certify the quality of qubits measurements beyond the von Neumann model and rank-one POVMs.
This takes essentially three steps. In the first step, the CHSH test is performed without the measurement to evaluate the quantity $\beta.$ In the second step, Bell tests are performed after the measurement to be certified is applied on the system, so as to evaluate the values $\cI_\theta$ and $\cI_\theta'$ (see appendix) for each classical output of the instrument. In the third step, the fidelity of the actual measurement is deduced from fidelities of the state before the measurement $F^i$ and of the post-measurement states $F_\theta^\ell$ using the formula given in Eq. \eqref{eq:FidLowBound}.
The robustness of our certification techniques together with the flexibility of our recipe make us confident that self-testing of quantum instruments could soon be demonstrated experimentally. For a more theoretical perspective, our results could be naturally extended to self-test only one Kraus operator within a family. This could be obtained by generalizing the 'heralded' fidelity defined in~\cite{Bancal18} to the current setting.\\

\begin{acknowledgments}
This work was supported by the Swiss National Science Foundation (SNSF), through the Grant PP00P2-179109 and 200021-175527. We also acknowledge the Army Research Laboratory Center for Distributed Quantum Information via the project SciNet.
\end{acknowledgments}

\appendix
\section{Appendix}

%%%%%%%%%%%%%%%%%
%\paragraph{\textbf{Appendix A -- Self-testing of a partially entangled two-qubit state}}
In the supplemental material we will prove the inequality in Eq.~(12) of the main text. To do so we have to lower bound the fidelity of an unknown state $\rho$  with respect to
\begin{equation}\label{eq: part state}
\ket{\phi_\theta^0} = \cos(\theta)\ket{00} + \sin(\theta)\ket{11}
\end{equation}
as a function of the expected value of the following Bell expression evaluated on the state $\rho$
\begin{align}
\cI_\theta &= \langle \frac14\left[\frac{A_0(B_0-B_1)}{\sin(b_\theta)}+\frac{\sin(2\theta)}{\cos(b_\theta)}A_1(B_0+B_1)\right.\nonumber\\
&  \ \ \ \ \ \ \ \ \ \ \ +\left.\cos(2\theta)\left(A_0 + \frac{B_0-B_1}{2\sin(b_\theta)}\right)\right]\rangle ,
\label{eq:SymBell}
\end{align}
where $b_\theta = \arctan\sqrt{\frac{1+\tfrac12 c_{2\theta}^2}{s_{2\theta}^2}}$. Here and in the rest of the appendix we use the short notation $c_\theta=\cos(\theta)$ and $s_\theta=\sin(\theta)$. The Bell expression Eq.~\eqref{eq:SymBell} has a quantum bound of $1$, achieved by measuring precisely the two-qubit state $\ket{\phi_\theta^0}$ with the observables
\begin{align}
A_0 &= \sigma_z \ , &  B_0 & = \cos(b_\theta)\sigma_x + \sin(b_\theta)\sigma_z, \\
A_1 &= \sigma_x \ , &  B_1 & = \cos(b_\theta)\sigma_x - \sin(b_\theta)\sigma_z .
\label{eq:PerfectSettings} 
\end{align}
We find that the local bound of Eq.~\eqref{eq:SymBell} achieved by the deterministic local strategy $\{A_0=A_1=B_0=1,B_1=-1\}$ is given by
\begin{align}
\cI_L = \frac14\left[c_{2\theta}+(2+c_{2\theta})\sqrt{\frac{7-c_{4\theta}}{5+c_{4\theta}}}\right] .
\label{eq:LocBoundSym}
\end{align}
This value is to be compared with the local bound of the well-known (normalized) tilted-CHSH 
\be
\cI_\theta^{(T)}=\left\langle \frac{\alpha_\theta A_0 + A_0 (B_0 + B_1) + A_1( B_0- B_1) }{\sqrt{8+2 \alpha_\theta^2}}\right\rangle
\ee
with $\alpha_\theta= \frac{2}{\sqrt{1+2 \tan^2(2\theta)}}$, also known to attain the quantum bound of $1$ for the partially entangled state of Eq.~\eqref{eq: part state} and well-chosen measurement settings, see Ref. [43, 44] of the main text. We find that the local bound of Eq.~\eqref{eq:LocBoundSym} is always higher than the local bound of the tilted-CHSH 
$\cI_L \geq \cI_L^{(T)}= \frac{2+\alpha_\theta}{\sqrt{8+2 \alpha_\theta^2}}$, meaning that the violation of the tilted-CHSH inequality is more robust to white noise than the violation of our new Bell inequality.
Nevertheless, our later studies will show that the new Bell operator allows for a more noise-tolerant state certification. The reason for this counter-intuitive result is the following: comparing the observed violation to the local bound only provides information on the distance between the target state (in our case $\ket{\phi_\theta^0}$) and deterministic strategies, given by parallel measurement setting $A_0=\pm A_1$ with $B_0=\pm B_1$ and product states $\ket{\psi_A}\ket{\psi_B}$ with $A_0\ket{\psi_A} = \pm \ket{\psi_A}$ and $B_0\ket{\psi_B} = \pm \ket{\psi_B}$. In self-testing, on the other hand, we need to bound the distance of the target state to arbitrary states for arbitrary measurement settings. It is then crucial that the violation worsens drastically when departing from the perfect settings.  

\subsection{Fidelity Bounds}

To derive lower bounds on the state fidelity from a Bell violation $\cI_\theta$ we use the tools presented in Ref.~[27] of the main text. There, it is shown that such a global lower bound can be obtained by solving a two-qubit problem.
More precisely, the state fidelity can be bounded by minimizing the quantity  
\be
O((\Lambda_\sA\otimes\Lambda_\sB^{o,\theta})[\rho],\prjct{\phi_\theta})
\ee
with $O(\rho,\sigma)=\tr(\sigma \rho)$, over all possible two-qubit states $\rho$ and all possible qubit observables $A_0,A_1,B_0,B_1$ (with eigenvalues $\pm 1$) that are compatible with the value $\cI_\theta$ of the Bell operator. Here,
\be
\Lambda_\sA, \Lambda_\sB^{o,\theta}: L(\mathds{C}^2) \rightarrow L(\mathds{C}^2 )
\ee
are the local extraction channels that can depend on the local observable $A_0$ with $A_1$ for $\Lambda_\sA$ and $B_0$ with $B_1$ for $\Lambda_\sB^{o,\theta}$. The first step, therefore, is to fix these extraction channels.

Before we do so, let us fix some notation for the local observables. Any qubit observable with eigenvalues $+1$ and $-1$ can be written as ${\bm n} \cdot {\bm \sigma}$ with $|{\bm n}|=1$. Furthermore, without loss of generality, we can set the local bases such that
\begin{align}
    A_0(a) &=\cos(a)H+\sin(a)V\nonumber \\ A_1(a) &=\cos(a)H-\sin(a)V \label{eq:A1}\\
    B_0(b) &=\cos(b)\sigma_x+\sin(b)\sigma_z \nonumber\\ B_1(b) &=\cos(b)\sigma_x-\sin(b)\sigma_z\label{eq:B1}
\end{align}
where $H=\tfrac1{\sqrt2}(\sigma_z+\sigma_x)$, $V = \tfrac1{\sqrt2}(\sigma_z-\sigma_x)$. Hence, in the minimization the pairs of observable $A_0$, $A_1$ and $B_1$, $B_2$ as well as the extraction channels $\Lambda_\sA$ and $\Lambda_\sB^{o,\theta}$ only depend on a single parameter $a$ and $b$ respectively.

\subsection{Extraction Channels}

The extraction channels we will use are adapted versions of the dephasing channels from Ref.~[41] of the main text. Note that since any local map can be seen as an isometry acting on the state plus the auxiliary degrees of freedom, these maps can also be understood as defining local isometries of particular interest. On Alice's side, the observables are maximally dephased if they are parallel or anti-parallel, and unchanged if they are orthogonal. More precisely, for $a$ being half the angle between Alice's observables in Eq.~\eqref{eq:A1} the dephasing acts according to
\begin{align}\label{eq: Lambda A}
\Lambda_a[\rho] := \frac{1+g(a)}2\rho + \frac{1-g(a)}2 \Gamma_a \rho\Gamma_a \ , 
\end{align}
where $g(a) = (1+\sqrt2)(\cos(a)+\sin(a)-1)$, and $\Gamma_a=H$ if $a\in[0,\frac\pi4]$ and $\Gamma_a=V$ if $a\in]\frac\pi4,\frac\pi2]$. Here and from now on, $\Lambda_a$ is the short notation of $\Lambda_\sA(a)$.

On Bob's side, the observables are also maximally dephased if they are parallel or anti-parallel, but here they are unchanged if half the angle between them equals $b_\theta$, where $b_\theta = \arctan\sqrt{\frac{1+\tfrac12 c_{2\theta}^2}{s_{2\theta}^2}}$ for the new inequality and $b_\theta=\arctan(\sin(2\theta))$ for the tilted-CHSH one. For $b$ denoting half the angle between Bob's observables in Eq.~\eqref{eq:B1} the dephasing channel on Bob's side is
\begin{align}
\Lambda_b[\rho] := \frac{1+g(t_\theta(b))}2\rho + \frac{1-g(t_\theta(b))}2 \Omega_b \rho\Omega_b \ , 
\end{align}
where
\begin{align}
t_\theta(b) &= \gamma_\theta^{-1} \ln\left(\frac{b-\delta_\theta}{\delta_\theta}\right),\\
\gamma_\theta &= \frac4\pi \ln\left(\frac{\frac\pi2 - b_\theta}{b_\theta}\right),\\
\delta_\theta &= \frac{b_\theta^2}{\pi^2-2b_\theta}.
\end{align}
For the observables of Bob, the dephasing happens in the direction $\Omega_b=\sigma_x$ if $b\in[0,b_\theta]$ and $\Omega_b=\sigma_z$ if $b\in]b_\theta,\frac\pi2]$. Again, we introduced the simplified notation $\Lambda_b=\Lambda_\sB^{o,\theta}(b)$.\\

Using these extraction channels and applying the numerical method presented Ref.~[27] of the main text, we find the values of $I_\theta^*$ such that the convex bound
\begin{align}\label{eq: overlap}
O((\Lambda_\sA\otimes\Lambda_\sB^{o,\theta})[\rho],\prjct{\phi_\theta^0}) \geq (1-c_\theta^2)\frac{\cI_\theta-\cI_\theta^*}{1-\cI_\theta^*}+c_\theta^2 
\end{align}
holds for our new inequality, as well as for the tilted-CHSH one (using then $\cI_\theta^{(T)}$ instead of $\cI_\theta$). Here $\cI_\theta$ is the observed Bell violation and $\cI_\theta^*$ is the non-trivial cutoff corresponding to the violation for which the fidelity matches the square of the largest Schmidt coefficient of $\ket{\phi_\theta^0}$. This means that $\cI_\theta^*$ is the relevant quantity for comparing the self-testing performance of Bell operators in device-independent tasks. Figure~4 in the main text depicts $\cI_\theta^*$ for both the tilted-CHSH and the new inequality.

\subsection{Extension to the other state}

\label{app: the other state}

We have just shown how the observed violation of a Bell inequalities $\cI_\theta$ gives a bound on the overlap of the state $\rho$ with the target state 
\begin{align}
\ket{\phi_\theta^0} = c_\theta \ket{00} + s_\theta \ket{11}
\end{align}
upon applying the extraction maps $\Lambda_a$ and $\Lambda_b$. We will now show the violation of another inequality $\cI_\theta'$, related to $\cI_\theta$ by a mere relabeling of some of the inputs and outputs, bounds the overlap of a state $\rho'$ with the other target state 
\begin{align}
\ket{\phi_\theta^1} = c_\theta \ket{11} + s_\theta \ket{00}.
\end{align}
The proof follows the line of Ref. [26] from the main text. Before we start, note that the two target states are related via $\ket{\phi_\theta^0}\overset{\sigma_x\otimes\sigma_x}{\longleftrightarrow}\ket{\phi_\theta^1}$. In the ideal case this transformation corresponds to a permutation of the outputs of $A_1\leftrightarrow -A_1$ and the exchange of $B_0$ and $B_1$. We will now show that this observation also holds in the non-ideal case.

Let us first have a closer look at Eq.~\eqref{eq: overlap}. As it holds for any state $\rho$, it can be expressed as the expectation value of the operator
\begin{align}
\Lambda_a\tot\Lambda_b \left[\prjct{\phi_\theta^0}\right] - s \cB(a,b) -\mu \id \geq 0,
\label{eq:OpIneq}
\end{align}
where $\cB(a,b)$ is the Bell operator obtained by choosing the settings~\eqref{eq:A1},~\eqref{eq:B1} in the Bell expression~\eqref{eq:SymBell}. Here we used the fact that the maps $\Lambda_{a(b)}$ are self-adjoint and $s= \frac{1-c_\theta^2}{1-\cI_\theta^*}$ and $\mu = \frac{c_\theta^2-\cI_\theta^*}{1-\cI_\theta^*}$. This operator inequality holds for all measurement angles $a$ and $b$. 

\begin{figure}
\centering
\includegraphics[width=0.9\columnwidth]{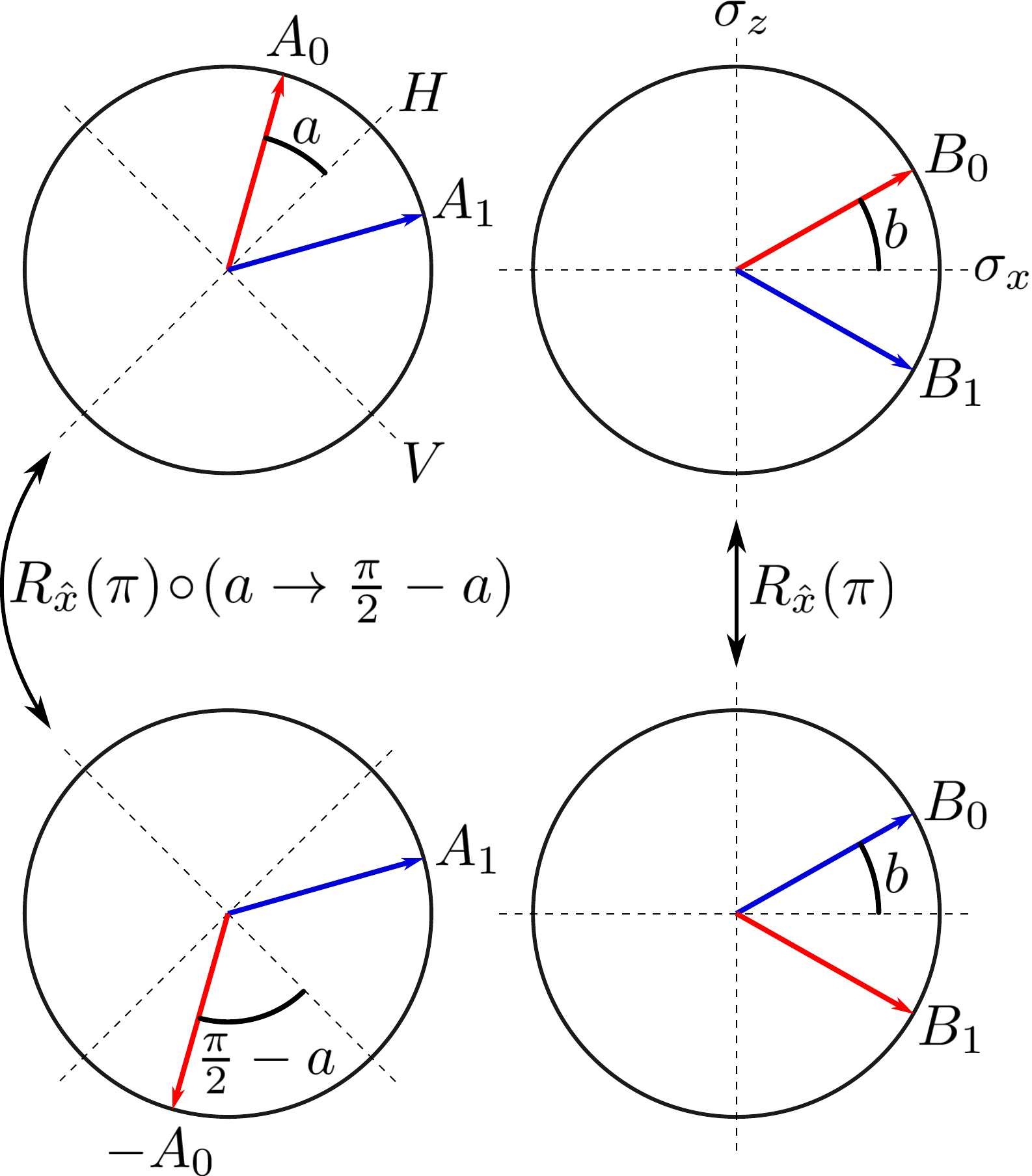}
\caption{The first row shows the settings present in the experiment, the left column belongs to Alice, the right one to Bob. In case the outcome of the generalized measurement is $0$, they just collect the data. If the outcome is $1$, Bob reinterprets his inputs, i.e. if he chose $B_0$ he saves the result as the outcome of $B_1$ and vice versa. Alice on the other hand collects her data applying $A_0 \to -A_0$. This transformation is sketched in the second row. The effect of this post-processing on the settings correspond to Alice applying a shift to her measurement angle followed by a rotation around $\hat{x}$ by $\pi$, i.e. $R_{\hat{x}}(\pi) \!\circ\! (a\to\frac\pi2-a)$, where $ R_{\hat{x}}(\pi)=e^{i \frac\pi2 \sigma_x}$. The observables of Bob are simply rotated by $R_{\hat{x}}(\pi)$.}
\label{fig:SettingsTrafo}
\end{figure}

Now consider a new Bell expression corresponding to an operator $\cB'$, obtained by exchanging the outputs of $A_1$ and exchanging the role of the measurements $B_0$ and $B_1$ in the previous expression. The operator $\cB'(a,b)$ can be obtained by applying the rotation $R := R_{\hat{x}}(\pi)=e^{i \frac\pi2 \sigma_x}$:
\begin{equation}
\cB'(a,b) = (R \otimes R)\cB'(\frac\pi2-a,b)(R^\dagger \otimes R^\dagger)
\end{equation}
as illustrated in Figure~\ref{fig:SettingsTrafo}.

Since Eq.~\eqref{eq:OpIneq} holds for all angles, it also holds for $a\to\frac\pi2-a$. Then we act with $(R\otimes R)$ and $(R^\dagger\otimes R^\dagger)$ from the left and right respectively. We arrive at
\begin{align}
(R\tot R)(\Lambda_{\tfrac\pi2-a}\!\tot\Lambda_b)\!\!\left[\prjct{\phi_\theta^0}\right]\!(R^\dagger\tot R^\dagger)& \label{eq:phi1Deriv1}\\
- s \cB'(a,b) -\mu \id &\geq 0 . \nonumber
\end{align}
The rotation on Bob's side commutes with the extraction channel $\Lambda_b$. This is easily verified by reminding ourselves that $\Omega_b \in \{\sigma_x,\sigma_z\}$ and therefore $R\Omega_b\rho\Omega_b R^\dagger = \Omega_b R\rho R^\dagger \Omega_b$.

On Alice's side, we realize that $R H R^\dagger = V$. Therefore $R \Gamma_{\tfrac\pi2-a} R^\dagger = \Gamma_a$. By also recalling that $g(\tfrac\pi2-a)=g(a)$, we find that $R \Lambda_{\tfrac\pi2-a}[\rho] R^\dagger = \Lambda_a[R\rho R^\dagger]$.

Combining everything and using $R\otimes R\ket{\phi_\theta^0}=\ket{\phi_\theta^1}$, Eq.~\eqref{eq:phi1Deriv1} is equivalent to 
\begin{align}
\Lambda_a\tot\Lambda_b \left[\prjct{\phi_\theta^1}\right] - s \cB'(a,b) -\mu \id \geq 0 ,
\label{eq:OpIneqPhi1}
\end{align}
which implies that with the same extraction channels and observables, and a new Bell test $\cI_\theta'$, one can device-independently self-test the second output state $\ket{\phi_\theta^1}$.
\\


\begin{thebibliography}{25}

\bibitem{Hensen15} B. Hensen, H. Bernien, A. E. Dréau, A. Reiserer, N. Kalb, M. S. Blok, J. Ruitenberg, R. F. L. Vermeulen, R. N. Schouten, C. Abellán, W. Amaya, V. Pruneri, M. W. Mitchell, M. Markham, D. J. Twitchen, D. Elkouss, S. Wehner, T. H. Taminiau, R. Hanson, \textit{Loophole-free Bell inequality violation using electron spins separated by 1.3 kilometres}, \href{https://doi.org/10.1038/nature15759}{Nature {\bf 526}, 682 (2015)}.

\bibitem{Shalm15} L. K. Shalm, E. Meyer-Scott, B. G. Christensen, P. Bierhorst, M. A. Wayne, M. J. Stevens, T. Gerrits, S. Glancy, D. R. Hamel, M. S. Allman, K. J. Coakley, S. D. Dyer, C. Hodge, A. E. Lita, V. B. Verma, C. Lambrocco, E. Tortorici, A. L. Migdall, Y. Zhang, D. R. Kumor, W. H. Farr, F. Marsili, M. D. Shaw, J. A. Stern, C. Abellán, W. Amaya, V. Pruneri, T. Jennewein, M. W. Mitchell, P. G. Kwiat, J. C. Bienfang, R. P. Mirin, E. Knill, S. W. Nam, \textit{Strong Loophole-Free Test of Local Realism}, \href{https://doi.org/10.1103/PhysRevLett.115.250402}{Phys. Rev. Lett. {\bf 115}, 250402 (2015)}.

\bibitem{Giustina15} 
M. Giustina, M. A. M. Versteegh, S. Wengerowsky, J. Handsteiner, A. Hochrainer, K. Phelan, F. Steinlechner, J. Kofler, J-Å. Larsson, C. Abellán, W. Amaya, V. Pruneri, M. W. Mitchell, J. Beyer, T. Gerrits, A. E. Lita, L. K. Shalm, S. W. Nam, T. Scheidl, R. Ursin, B. Wittmann, A. Zeilinger, \textit{Significant-Loophole-Free Test of Bell’s Theorem with Entangled Photons}, \href{https://doi.org/10.1103/PhysRevLett.115.250401}{Phys. Rev. Lett. {\bf 115}, 250401 (2015)}.

\bibitem{Rosenfeld17} W. Rosenfeld, D. Burchardt, R. Garthoff, K. Redeker, N. Ortegel, M. Rau, H. Weinfurter, \textit{Event-Ready Bell Test Using Entangled Atoms Simultaneously Closing Detection and Locality Loopholes}, \href{https://doi.org/10.1103/PhysRevLett.119.010402}{Phys. Rev. Lett. {\bf 119}, 010402 (2017)}.

\bibitem{Bell64} J.S. Bell, \textit{On the Einstein Podolsky Rosen paradox}, \href{https://doi.org/10.1103/PhysicsPhysiqueFizika.1.195}{Physics {\bf 1}, 195 (1964)}.

\bibitem{Colbeck09} R. Colbeck, \textit{Quantum And Relativistic Protocols For Secure Multi-Party Computation}, \href{https://arxiv.org/abs/0911.3814}{Ph.D. thesis, (2009)}.

\bibitem{Pironio10} S. Pironio, A. Acín, S. Massar, A. Boyer de la Giroday, D. N. Matsukevich, P. Maunz, S. Olmschenk, D. Hayes, L. Luo, T. A. Manning, C. Monroe, \textit{Random numbers certified by Bell’s theorem}, \href{https://doi.org/10.1038/nature09008}{Nature {\bf 464}, 1021 (2010)}.

\bibitem{Christensen13} B. G. Christensen, K. T. McCusker, J. B. Altepeter, B. Calkins, T. Gerrits, A. E. Lita, A. Miller, L. K. Shalm, Y. Zhang, S. W. Nam, N. Brunner, C. C. W. Lim, N. Gisin, and P. G. Kwiat, \textit{Detection-Loophole-Free Test of Quantum Nonlocality, and Applications}, \href{https://doi.org/10.1103/PhysRevLett.111.130406}{Phys. Rev. Lett. {\bf 111}, 130406 (2013)}.

\bibitem{Yang18} Y. Liu, X. Yuan, M-H. Li, W. Zhang, Q. Zhao, J. Zhong, Y. Cao, Y-H. Li, L-K. Chen, H. Li, T. Peng, Y-A. Chen, C-Z. Peng, S-C. Shi, Z. Wang, L. You, X. Ma, J. Fan, Q. Zhang, J-W. Pan, \textit{High-Speed Device-Independent Quantum Random Number Generation without a Detection Loophole}, \href{https://doi.org/10.1103/PhysRevLett.120.010503}{ Phys. Rev. Lett. {\bf 120}, 010503 (2018)}.

\bibitem{Yang18bis} Y. Liu, Q. Zhao, M-H. Li, J-Y. Guan, Y. Zhang, B. Bai, W. Zhang, W-Z. Liu, C. Wu, X. Yuan, H. Li, W. J. Munro, Z. Wang, L. You, J. Zhang, X. Ma, J. Fan, Q. Zhang, J-W. Pan, \textit{Device-independent quantum random-number generation}, \href{https://doi.org/10.1038/s41586-018-0559-3}{Nature {\bf 562}, 548 (2018)}.

\bibitem{Bierhorst18} P. Bierhorst, E. Knill, S. Glancy, Y. Zhang, A. Mink, S. Jordan, A. Rommal, Y-K. Liu, B. Christensen, S. W. Nam, M. J. Stevens, L. K. Shalm, \textit{Experimentally Generated Randomness Certified by the Impossibility of Superluminal Signals}, \href{https://doi.org/10.1038/s41586-018-0019-0}{Nature {\bf 556}, 223 (2018)}.

\bibitem{Shen18} L. Shen, J. Lee, L. P. Thinh, J-D. Bancal, A. Cerè, A. Lamas-Linares, A. Lita, T. Gerrits, S. W. Nam, V. Scarani, C. Kurtsiefer \textit{Randomness Extraction from Bell Violation with Continuous Parametric Down-Conversion}, \href{https://doi.org/10.1103/PhysRevLett.121.150402}{Phys. Rev. Lett. {\bf 121}, 150402 (2018)}.

\bibitem{CHSH69} J. F. Clauser, M. A. Horne, A. Shimony, R.A. Holt, \textit{Proposed Experiment to Test Local Hidden-Variable Theories}, \href{https://doi.org/10.1103/PhysRevLett.23.880}{Phys. Rev. Lett. {\bf 23}, 880 (1969)}.

\bibitem{Popescu92} S. Popescu, D. Rohrlich, \textit{Which states violate Bell's inequality maximally?}, \href{https://doi.org/10.1016/0375-9601(92)90819-8}{Phys. Lett. A {\bf 169}, 411 (1992)}.

\bibitem{McKague12} M. McKague, T. H. Yang, V. Scarani, \textit{Robust self-testing of the singlet}, \href{https://doi.org/10.1088/1751-8113/45/45/455304}{J. Phys. A: Math. Theor. {\bf 45}, 455304 (2012)}.

\bibitem{Mayers04} D. Mayers, A. Yao, \textit{Quantum Cryptography with Imperfect Apparatus}, \href{https://arxiv.org/abs/quant-ph/9809039}{Proceedings of the 39th IEEE Conference on Foundations of Computer Science, 1998, page 503}, see also \textit{Self testing quantum apparatus}, \href{https://doi.org/10.26421/QIC4.4}{Quant. Inf. Comput. {\bf 4}, 273 (2004)}.

\bibitem{McKague13} M. McKague, \textit{Interactive Proofs for BQP via Self-Tested Graph States}, \href{https://doi.org/10.4086/toc.2016.v012a003}{Theory of Computing, {\bf 12}, 3 (2016)}.

\bibitem{Coladangelo16} A. Coladangelo, K.T. Goh, V. Scarani, \textit{All Pure Bipartite Entangled States can be Self-Tested}, \href{https://doi.org/10.1038/ncomms15485}{Nat. Comm. {\bf 8}, 15485 (2017)}.

\bibitem{Wu14} X. Wu, Y. Cai, T. H. Yang, H. Nguyen Le, J-D. Bancal, V. Scarani, \textit{Robust self-testing of the three-qubit W state}, \href{https://doi.org/10.1103/PhysRevA.90.042339}{Phys. Rev. A {\bf 90}, 042339 (2014)}.

\bibitem{Bancal15} J-D. Bancal, M. Navascués, V. Scarani, T. Vértesi, T. H. Yang, \textit{Physical characterization of quantum devices from nonlocal correlations}, \href{https://doi.org/10.1103/PhysRevA.91.022115}{Phys. Rev. A {\bf 91}, 022115 (2015)}.

\bibitem{Chen16} S-L. Chen, C. Budroni, Y-C. Liang, Y-N. Chen, \textit{Natural Framework for Device-Independent Quantification of Quantum Steerability, Measurement Incompatibility, and Self-Testing}, \href{https://doi.org/10.1103/PhysRevLett.116.240401}{Phys. Rev. Lett. {\bf 116}, 240401 (2016)}.

\bibitem{Cavalcanti16} D. Cavalcanti, P. Skrzypczyk, \textit{Quantitative relations between measurement incompatibility, quantum steering, and nonlocality}, \href{https://doi.org/10.1103/PhysRevA.93.052112}{Phys. Rev. A {\bf 93}, 052112 (2016)}.

\bibitem{Kaniewski17} J. Kaniewski, \textit{Self-testing of binary observables based on commutation}, \href{https://doi.org/10.1103/PhysRevA.95.062323}{Phys. Rev. A {\bf 95}, 062323 (2017)}.

\bibitem{Bowles18} J. Bowles, I. Šupić, D. Cavalcanti, A. Acín, \textit{Self-testing of Pauli observables for device-independent entanglement certification}, \href{https://doi.org/10.1103/PhysRevA.98.062336}{Phys. Rev. A {\bf 98}, 042336 (2018)}.

\bibitem{Renou18} M-O. Renou, J. Kaniewski, N. Brunner, \textit{Self-Testing Entangled Measurements in Quantum Networks}, \href{https://doi.org/10.1103/PhysRevLett.121.250507}{Phys. Rev. Lett. {\bf 121}, 250507 (2018)}.

\bibitem{Bancal18} J-D. Bancal, N. Sangouard, P. Sekatski, \textit{Noise-Resistant Device-Independent Certification of Bell State Measurements}, \href{https://doi.org/10.1103/PhysRevLett.121.250506}{Phys. Rev. Lett. {\bf 121}, 250506 (2018)}.

\bibitem{Sekatski18} P. Sekatski, J-D. Bancal, S. Wagner, N. Sangouard, \textit{Certifying the Building Blocks of Quantum Computers from Bell’s Theorem}, \href{https://doi.org/10.1103/PhysRevLett.121.180505}{Phys. Rev. Lett. {\bf 121}, 180505 (2018)}.

\bibitem{Gomez16} E. S. Gómez, S. Gómez, P. González, G. Cañas, J. F. Barra, A. Delgado, G. B. Xavier, A. Cabello, M. Kleinmann, T. Vértesi, G. Lima, \textit{Device-Independent Certification of a Nonprojective Qubit Measurement}, \href{https://doi.org/10.1103/PhysRevLett.117.260401}{Phys. Rev. Lett. {\bf 117}, 260401 (2016)}.

\bibitem{Smania18} M. Smania, P. Mironowicz, M. Nawareg, M. Pawłowski, A. Cabello, M. Bourennane, \textit{Experimental certification of an informationally complete quantum measurement in a device-independent protocol}, \href{https://doi.org/10.1364/OPTICA.377959}{Optica {\bf 7},123-128 (2020)}.

\bibitem{davies70} E. B. Davies, J. T. Lewis, \textit{An operational approach to quantum probability}, \href{https://doi.org/10.1007/BF01647093}{Comm. Math. Phys. {\bf 17}, 239 (1970)}.

\bibitem{Gross18} J. A. Gross, C. M. Caves, G. J. Milburn, J. Combes, \textit{Qubit models of weak continuous measurements: markovian conditional and open-system dynamics}, \href{https://doi.org/10.1088/2058-9565/aaa39f}{Quantum Sci. Technol. {\bf 3}, 024005 (2018)}.

\bibitem{Silva15} R. Silva, N. Gisin, Y. Guryanova, S. Popescu, \textit{Multiple Observers Can Share the Nonlocality of Half of an Entangled Pair by Using Optimal Weak Measurements}, \href{https://doi.org/10.1103/PhysRevLett.114.250401}{Phys. Rev. Lett. {\bf 114}, 250401 (2015)}.

\bibitem{Curchod17} F. J. Curchod, M. Johansson, R. Augusiak, M. J. Hoban, P. Wittek, A. Acín \textit{Unbounded randomness certification using sequences of measurements}, \href{https://doi.org/10.1103/PhysRevA.95.020102}{Phys. Rev. A {\bf 95}, 020102 (2017)}.

\bibitem{Curchod18} F. J. Curchod, M. Johansson, R. Augusiak, M. J. Hoban, P. Wittek, A. Acín, \textit{A Single Entangled System Is an Unbounded Source of Nonlocal Correlations and of Certified Random Numbers}, \href{https://doi.org/10.4230/LIPIcs.TQC.2017.1}{12$^\text{th}$ Conference on the Theory of Quantum Computation, Communication and Cryptography (TQC 2017), Leibniz International Proceedings in Informatics (LIPIcs) \bf{73}, 1:1 (2018)}.

\bibitem{Coyle18} B. Coyle, M.J. Hoban, E. Kashefi, \textit{One-Sided Device-Independent Certification of Unbounded Random Numbers}, \href{https://doi.org/10.4204/EPTCS.273.2}{EPTCS 273, 14 (2018)}.

\bibitem{Banaszek01} K. Banaszek, I. Devetak, \textit{Fidelity trade-off for finite ensembles of identically prepared qubits}, \href{https://doi.org/10.1103/PhysRevA.64.052307}{Phys. Rev. A {\bf 64}, 052307 (2001)}.

\bibitem{Choi75} M-D. Choi, \textit{Completely positive linear maps on complex matrices}, \href{https://doi.org/10.1016/0024-3795(75)90075-0}{Linear Algebra Appl., {\bf 10}, 285 (1975)}.

\bibitem{Jamiolkowski72} A. Jamiołkowski, \textit{Linear transformations which preserve trace and positive semidefiniteness of operators}, \href{https://doi.org/10.1016/0034-4877(72)90011-0}{Rep. Math. Phys. {\bf 3}, 275 (1972)}.

\bibitem{Raginsky01} M. Raginsky, \textit{A fidelity measure for quantum channels}, \href{https://doi.org/10.1016/S0375-9601(01)00640-5}{Phys. Lett. A {\bf 290}, 11 (2001)}.

\bibitem{Belavkin05} V. P. Belavkin, G. M. D’Ariano, M. Raginsky, \textit{Operational distance and fidelity for quantum channels}, \href{https://doi.org/10.1063/1.1904510}{J. Math. Phys. {\bf 46}, 062106 (2005)}.

\bibitem{Kaniewski16} J. Kaniewski, \textit{Analytic and Nearly Optimal Self-Testing Bounds for the Clauser-Horne-Shimony-Holt and Mermin Inequalities}, \href{https://doi.org/10.1103/PhysRevLett.117.070402}{Phys. Rev. Lett. {\bf 117}, 070402 (2016)}.

\bibitem{inprep} P. Sekatski et al., in preparation.

\bibitem{Acin12} A.~Acín, S.~Massar, S.~Pironio, \textit{Randomness versus Nonlocality and Entanglement}, \href{https://doi.org/10.1103/PhysRevLett.108.100402}{Phys. Rev. Lett, {\bf 108}, 100402 (2012)}.

\bibitem{Bamps15} C. Bamps, S.~Pironio, \textit{Sum-of-squares decompositions for a family of Clauser-Horne-Shimony-Holt-like inequalities and their application to self-testing}, \href{https://doi.org/10.1103/PhysRevA.91.052111}{ Phys. Rev. A {\bf 91}, 052111 (2015)}.

\bibitem{inprepa} T. Coopmans, J. Kaniewski, C. Schaffner, \textit{Robust self-testing of two-qubit states}, \href{https://doi.org/10.1103/PhysRevA.99.052123}{Phys. Rev. A {\bf 99}, 052123 (2019)}.

\end{thebibliography}
\end{document}